# Formation of Porous Gas Hydrates

Andrey N. Salamatin [(1,2)] and Werner F. Kuhs [(2)*]

[(1)] Dept. of Applied Mathematics, Kazan State University, Kazan 420008, Russia;
[(2)] GZG Abt. Kristallographie, Georg-August-Universität Göttingen, Göttingen 37077, Germany

Gas hydrates grown at gas-ice interfaces are examined by electron microscopy and found to have a submicron porous texture. Permeability of the intervening hydrate layers provides the connection between the two counterparts (gas and water molecules) of the clathration reaction and makes further hydrate formation possible. The study is focused on phenomenological description of principal stages and rate-limiting processes that control the kinetics of the porous gas hydrate crystal growth from ice powders. Although the detailed physical mechanisms involved in the porous hydrate formation still are not fully understood, the initial stage of hydrate film spreading over the ice surface should be distinguished from the subsequent stage which is presumably limited by the clathration reaction at the ice-hydrate interface and develops after the ice grain coating is finished. The model reveals a time dependence of the reaction degree essentially different from that when the rate-limiting step of the hydrate formation at the second stage is the gas and water transport (diffusion) through the hydrate shells surrounding the shrinking ice cores. The theory is aimed at the interpretation of experimental data on the hydrate growth kinetics.

## 1 Introduction

Gas clathrate hydrates exist at certain thermodynamic conditions as solid solutions of the gas (guest) molecules in the metastable crystalline lattice of water (host) molecules. The guest molecules fill non-stoichiometrically the lattice cavities (cages), stabilising the compound (Van der Waals and Platteeuw, 1959; Sloan, 1997). Two main crystallographic structures of gas hydrates, the Stackelberg Structure I and II, are distinguished both consisting of the two types of cavities, small and large cages, that can be occupied by the guest molecules.

Since the 1950ies, different hydrate crystals have been produced and studied in laboratories. Still, some physical properties of gas hydrates and the kinetics of their formation (or dissociation) are neither well known nor properly understood. One of the new and most intriguing findings is that, at least in certain cases, some clathrate-hydrate crystals grow with a sponge-like porous microstructure. Using cryo field-emission scanning electron microscopy (FE-SEM), first direct observations of such meso- to macro-porous gas hydrates were made (Kuhs et al., 2000). The typical diameter of the pores was established as 100-400 nm for $CH_4$, Ar and $N_2$ hydrates and 20-100 nm for $CO_2$ hydrate, while the ice Ih powder used as starting material had a grain size of a few hundred μm. Optical microscopic investigations by Kobayashi and others (2001) also indicated a coarse and possibly porous microstructure of the hydrate films grown at liquid-liquid interfaces, especially in contact with flowing water. Rather interestingly, there are evidences that some natural gas hydrates from the ocean sea floor also exhibit sub-micron porosity (Bohrmann et al., these proceedings). Based on experimental studies (Aya et al., 1992; Uchida and Kawabata, 1995; Sugaya and Mori, 1996) of $CO_2$ and fluorocarbon hydrate growth at liquid-liquid interfaces, Mori and Mochizuki (1997) and Mori (1998) had already proposed a porous microstructure of the hydrate layers intervening the two liquid phases and suggested a phenomenological capillary permeation model of water transport across the films. Although general physical concepts of this phenomenon in different situations may be quite similar, we still do not have sufficient data to develop a unified theoretical approach to its modelling (Mori, 1998). For this reason, our study is confined to the particular thermodynamic conditions of clathrate-hydrate formation on ice-grain surfaces in a single-gas atmosphere under relatively high pressures exceeding the dissociation pressure at fixed temperatures below the quadruple point.

In accordance with experimental observations (Uchida et al., 1992, 1994; Stern et al., 1998; Kuhs et al, 2000), a thin gas-hydrate film rapidly spreads over the ice surface at the initial (first) stage of the ice-to-hydrate conversion. After this, the only possibility to maintain the clathration reaction is the penetration of the gaseous phase through the intervening hydrate layer to the ice-hydrate interface and/or water molecules to the outer hydrate interface with the ambient gas. The diffusion-limited clathrate-crystal growth on ice surface at the subsequent (second) stage was considered and simulated by Fujii and Kondo (1974), Hondoh and Uchida (1992), Salamatin et al. (1998), Henning et al. (2000), Takeya et al. (2000). In the case of the porous gas-hydrate layer, the gas and water mass transport through it is expected to become much easier, and the clathration reaction itself is thought to be the rate-limiting step of the hydrate formation process. As has been repeatedly invoked by many investigators (e.g. Hwang et al., 1990; Sloan and Fleyfel, 1991; Henning et al., 2000; Takeya et al., 2000), the quasi-liquid layer (QLL) of water molecules at the ice-hydrate interface may play an important role in the gas distribution (diffusion) over the phase boundary and may be the key attribute of the reaction although a definite proof is still missing.

The principal goal of our paper is to summarise available FE-SEM images and other results of laboratory studies of porous clathrate hydrates in order to develop a phenomenological two-staged model of the hydrate growth from ice powders as a tool for the kinetics data interpretation. Assuming here that the transformation of water and gas molecules into a hydrate crystal (taking place at the ice core – hydrate interface) is the rate-limiting step at the second stage, we still can expect the onset of a further third stage of the clathration process limited by the gas and water mass transport through the hydrate shells when a highly consolidated ice-hydrate structure develops. This final period will be dealt with in a separate paper.

---

* Corresponding author. E-mail: wf.kuhs@geo.uni-goettingen.de.

## 2 Experimental Study of Porous Gas Hydrates

Laboratory studies and the direct FE-SEM observations of the porous gas hydrates as well as general concepts allow us to formulate some more or less definite statements that constrain possible theoretical considerations related to our neutron diffraction experiments (Staykova et al., these proceedings).

1. The ice Ih powder was prepared of spherical grains of a few hundred of μm in diameter. Hydrate crystallites grown on the ice surface are rather big (a few μm) and each of them contains pores in its interior. The typical diameter of the pores estimated in FE-SEM measurements in agreement with Kuhs et al. (2000) is 100-400 nm for $CH_4$, Ar and $N_2$ hydrates and 20-100 nm for $CO_2$ hydrate.

2. The initial macro-porosity of the ice powder samples is determined as 15-25%, and the specific surface of ice is close to that of the grains with minimum sintering. The meso-porosity of hydrate phase is estimated to be ~10-20%, depending on thermodynamic conditions of the clathration reaction. The adsorption isotherm measurements show that the meso-pores are open.

3. The typical time scales of the heat and mass transfer processes in the ice powder samples are small (5-10 min). The temperature in the pressure cell in thermostart is practically uniform, and all substances and energy are rapidly redistributed within the solid phases and the macro-pore space.

4. The starting stage of the ice grain surface coverage by the clathrate film is clearly distinguished from the subsequent stage of the hydrate shell growth around the shrinking isolated ice cores.

5. The diffraction data suggest a good crystallinity of the hydrate crystals, indicating a strainless "inward" growth of hydrate without any appreciable deformation. This may occur only when the hydrate shell does not move with respect to ice. The latter conclusion is also confirmed by the FE-SEM pictures.

6. The density of the crystalline hydrate lattice of both types I and II is noticeably less than that of the ice. Thus, the excess water molecules must be partly "extracted" from the ice-hydrate contact area to vacate additional space for the newly formed clathrate and to develop the pores. This water (~20-30%) must be transported through the pore channels toward the outer hydrate surface to finally react with the ambient gas.

7. Total occupancy of the gas hydrate formed in non-equilibrium thermodynamic conditions is measured to be close to that in equilibrium with the ambient gas rather than in equilibrium with the ice. This is an indication that the surface premelting effects (e.g. Dash and others, 1995; Wettlaufer, 1999) may facilitate the gas and water transport along the gas- and ice-hydrate interfaces and prevent the pore closure. However, the clathration rates remain substantial even at temperatures as low as 230-240 K far beyond the range of the quasi-liquid layer existence, and the premelting is not the only mechanism that is involved in the gas-hydrate formation process. As mentioned in the introduction, the importance of the QLL was earlier emphasised in (Hwang et al., 1990; Sloan and Fleyfel, 1991; Henning et al., 2000; Takeya et al., 2000) but its domination might also be questioned because the obtained results do not reveal the expected strong increase in the growth rates on approaching the ice melting point.

8. The higher is the mean size of ice grains the less complete is the degree of transformation of ice to hydrate phase when the clathration reaction stops, indicating that the thick hydrate layers lose their permeability or/and the macro-pores become closed due to the sample compaction.

Further experiments and realistic data on the properties of the gas-ice-hydrate system should reveal a more distinct picture of the porous clathrate hydrate growth on ice surfaces. The next step of the study is to develop phenomenological mathematical description of the process at different stages of the clathration as an instrument for the kinetics data interpretation and stage identification.

## 3 Models for Hydrate-Formation Stages Limited by Clathration Reaction

In our consideration the current geometry of the ice powder structure is described in terms of the mean ice grain (core) radius $r_i$ and the specific surface area of ice grains per mole of water molecules $S_i$ (with $r_{i0}$ and $S_{i0}$ for their initial values); $\rho_i$ is the molar density of ice. Due to annealing, ice grains in a sample are often connected by necks. Nevertheless, hereinafter we assume that $S_{i0}$ is equal to the sum of the spherical grain surfaces and by definition,

$$S_i = S_{i0}(1-\alpha)^{2/3}, \quad S_{i0} = 3/(r_{i0}\rho_i), \qquad (1)$$

where $\alpha$ designates the total degree of the reaction (the mole fraction of ice converted to hydrate phase) and is the principal characteristic of the hydrate formation process developing in time $t$.

Furthermore, we introduce $\alpha_S$ to describe the fraction of the ice surface covered with hydrate. We also use $\omega_S$ and $\omega_V$ to denote the rates of the ice surface coating and the ice-to-hydrate transformation. The former quantity can be defined as the fraction of the open (exposed to the ambient gas) ice surface which becomes covered by the initial hydrate film during a unit time period, while the latter one is the number of ice moles transformed to hydrate phase per unit of time on unit ice surface area after its coating. Actually, $\omega_V$ describes two simultaneous reactions at the external boundary between the hydrate layer and gaseous phase and at the ice-hydrate interface of which the second one is considered as the rate-limiting step.

We conventionally write

$$\omega_J = k_J \left(\frac{p - p_d}{p_d}\right)^\beta \exp\left[\frac{Q_J}{R}\left(\frac{1}{T_*} - \frac{1}{T}\right)\right]. \qquad (2)$$

Here $p$ and $T$ are the gaseous phase pressure (fugacity) and the thermostart temperature, respectively; $k_J$ and $Q_J$ are the clathration rate constant at the reference temperature $T_*$ and the activation energy of the $J$-type reaction, $J = S, V$, $p_d$ is the hydrate dissociation pressure (gas fugacity); $R$ is the gas constant. The driving force of the reaction is the supersaturation of the gas-ice-hydrate system. It is represented in Eq. (2) by the reduced pressure factor raised to power $\beta$ which approximates general expressions for the reaction rates (e.g. Istomin and Yakushev, 1992; Markov, 1995) and accounts for their non-linearity.

**3.1 Initial Stage of Hydrate Film Spreading over the Ice Grain Surface.** In the beginning, the hydrate formation from the ice powder is controlled by the rate of the clathrate film spreading over the ice surface directly exposed to the ambient gas phase. At this stage $\alpha_S$ is determined by the well understandable equation

$$\frac{d\alpha_S}{dt} = \omega_S (1 - \alpha_S) \qquad (3a)$$

from which we can directly deduce

$$\alpha_S = 1 - e^{-\omega_S t} . \qquad (3b)$$

Let us also introduce the thickness of the ice layer, $\delta_0$, converted to the initial hydrate film in the coating process. This parameter is small compared to the mean grain size, but the clathration rate due to the hydrate film formation $\rho_i \delta_0 \omega_S$ is assumed to be much higher than $\omega_V$ on the hydrate-coated ice surface. Thus, the ice surface area remains practically constant ($S_i \approx S_{i0}$) during the initial stage of the clathration and the rate of the reaction degree growth in the sample can be expressed from Eqs. (1)-(3) as

$$\frac{d\alpha}{dt} = S_{i0} \left[ \rho_i \delta_0 \omega_S e^{-\omega_S t} + \omega_V \left(1 - e^{-\omega_S t}\right) \right] .$$

Finally, after the integration one obtains

$$\alpha / 3 = A \left(1 - e^{-\omega_S t}\right) + Bt , \qquad (4)$$

where

$$A = \frac{\delta_0}{r_{i0}} \left(1 - \frac{\omega_V}{\rho_i \delta_0 \omega_S}\right), \quad B = \frac{\omega_V}{r_{i0} \rho_i} .$$

It is worth noting that $B$ is the slope of the reaction degree growth at the end of the first stage which, as mentioned above, is expected to be small relative to $\omega_S$ and $A\omega_S$. This means, in particular, that for invariable $\delta_0$ the increase in $\alpha$ during the hydrate film formation on the ice grain surface does not depend noticeably on the reaction rates and thermodynamic conditions, especially if the activation energies $Q_S$ and $Q_V$ do not differ much. It is also important that coefficients in Eq. (4) are inversely proportional to the initial grain size $r_{i0}$.

**3.2 Hydrate Growth after Completion of the Ice Grain Coating.** Here we assume that the clathration reaction still remains the rate-limiting step of the gas hydrate growth around ice cores after the completion of the ice grain coverage by the primary hydrate film.

Hence, starting from the end of the initial stage, for $t > t^* \sim \omega_S^{-1}$ we have

$$\frac{d\alpha}{dt} = S_i \omega_V .$$

The substitution of Eq. (1) into the latter equation yields

$$\frac{d\alpha}{dt} = 3B(1 - \alpha)^{2/3}$$

and after integration we obtain

$$(1 - \alpha)^{1/3} = B(t^* - t) + (1 - \alpha^*)^{1/3} , \qquad (5)$$

where $\alpha^*$ is the degree of the reaction at the stage-transition time $t^*$.

It should be emphasised that Eq. (5) prescribes a linear dependence of $(1-\alpha)^{1/3}$ on time in stead of $(t - t^*)^{1/2}$ in the analogous predictive formula based on the diffusion model of Fujii and Kondo (1974) and used by Henning et al. (2000) for their data interpretation.

To determine parameters $\alpha^*$ and $t^*$, Eqs. (4) and (5) must be matched by equating their respective far-field and near-field expansions. For small $\alpha \sim \alpha^* \ll 1$ Eq. (4) can be represented as

$$(1 - \alpha)^{1/3} = 1 - A\left(1 - e^{-\omega_S t}\right) - Bt \qquad (6a)$$

and coincides with Eq. (5) for $\omega_S t \gg 1$ if $\alpha^*$ and $t^*$ are fixed so that

$$(1 - \alpha)^{1/3} = 1 - A - Bt . \qquad (6b)$$

Consequently, it becomes obvious that Eq. (6a) should be regarded as a combined uniform asymptotic solution applicable at any time when the clathration reaction is the rate-limiting step of the gas-hydrate formation process, whereas Eq. (6b) is the asymptotic presentation of Eq. (6a) for large $t$.

**3.3 General Solution.** If the rates of the ice grain coating and the hydrate layer growth are commensurable the ice core surface area $S_i$ noticeably decreases during the initial stage. In this case a unified general mathematical description of the porous gas hydrate formation becomes more complicated.

At any time period from $\tau$ to $\tau + d\tau$ a new portion of ice-grain surface $d\alpha_S$ becomes covered by the hydrate film, and from Eq. (3b) we have

$$d\alpha_S = \omega_S e^{-\omega_S \tau} d\tau .$$

The thickness of ice layer used for the initial film construction is $\delta_0$. After coating, the ice-hydrate interface moves toward an ice-core centre at the constant velocity $\omega_V / \rho_i$, and at any time $t > \tau$ the distance form the centre is

$$\Delta = r_{i0} - \delta_0 - \omega_V (t - \tau) / \rho_i .$$

Correspondingly, the increment of $\alpha$ can be expressed as

$$d\alpha = \left[1 - \left(\Delta / r_{i0}\right)^3\right] d\alpha_S ,$$

and the exact solution of the problem takes the integral form:

$$\alpha = \omega_S \int_0^t \left[1 - \left(1 - \frac{\delta_0}{r_{i0}} - B(t-\tau)\right)^3\right] e^{-\omega_S \tau} d\tau . \quad (7)$$

Its asymptotic expansion to terms of $O(B/\omega_S)^2$-order of magnitude

$$\alpha \approx 1 - (1 - A - Bt)^3 - \left[1 - (1-A)^3\right] e^{-\omega_S t}$$

practically reproduces Eq. (6) for small $A << 1$.

Most likely, the second stage limited by the clathration reaction or/and by the gas transport along the ice-hydrate interface does not continue until the end of the ice-to-hydrate transformation and is finally replaced by the third stage limited by the gas transfer through the hydrate shells surrounding the shrinking ice cores. This problem will be considered elsewhere.

## 4 Discussion

The above analysis presents a number of theoretical relations to study the process of the porous gas hydrate growth from ice powders. Eqs. (6) are the most simple ones among them, although they seem to be sufficiently accurate for the kinetics data interpretation. Eq. (6b) requires that the plot of $r_i/r_{i0} = (1-\alpha)^{1/3}$ against time $t$ during the second stage of the hydrate formation limited by the clathration reaction should be a straight line with slope $B$ and intercept $1-A$. Parameter $\omega_S$ in Eq. (6a) is determined by the curvature of the reaction degree graph at the initial stage of the ice surface coating. Parameters $\delta_0$ and $\omega_V$ can easily be expressed via $A$ and $B$ (see Eqs. (4)), provided that $r_{i0}$ (and $\rho_i$) is known. The $B$-slopes and $\omega_S$ measured at different temperatures directly result in estimates of the activation energies $Q_V$ and $Q_S$ from Eqs. (2).

It is important to note that in case of the hydrate formation process limited by the gas and water transport through the hydrate shells surrounding the shrinking ice cores the time behaviour of the quantity $(1-\alpha)^{1/3}$ becomes non-linear. This diffusion stage may follow the reaction-limited one or, at least in principle, may directly proceed from the initial stage, especially if the gas hydrates are not porous. Correspondingly, at small $\alpha$, in the beginning of the hydrate growth described by the simplified diffusion theory of Fujii and Kondo (1974), the relative ice core radius $r_i/r_{i0}$ is proportional to $(t-t^*)^{1/2}$. More sophisticated models (Salamatin et al., 1998; Takeya et al., 2000) predict even higher non-linearity due to the decrease in the ice-core surface $S_i$. Still, they do not take into account the sample compaction and the reduction of the macro-pore surface in the course of the ice-to-hydrate transformation. These effects additionally suppress the gas and water fluxes through the hydrate shells to and from the ice cores and will be considered elsewhere.

Another peculiarity of the diffusion-limited conversion of ice powders to clathrate hydrates is that the hydrate-growth rates in this case are inversely proportional to $r_{i0}^2$ (e.g. Salamatin et al., 1998; Takeya et al., 2000), being in contrast to Eqs. (4) and (6) with $A$ and $B$ inversely proportional to $r_{i0}$.

Thus, the stages controlled by different rate-limiting steps (clathration reaction or gas and water transport through the hydrate shells) can be distinguished one from another. This may also help to recognise the formation of porous gas hydrates in the kinetics data analysis.

In the companion paper (Staykova et al., these proceedings), neutron diffraction was used to study the formation of porous methane hydrates from deuterated ice powder samples. The first experiments show a promising agreement with the presented phenomenological approach.

A persisting problem is that we still do not have sufficient knowledge about mechanisms of the porous gas hydrate growth. Much more data are needed to specify the details of the thermodynamics and to achieve a necessary physical understanding of the kinetics of the non-equilibrium conversion of ice to clathrate hydrate.

## 5 Conclusion

A cryo field-emission scanning electron microscopy technique has been used to study porous gas hydrates grown on spherical ice grains. Although the detailed mechanisms of the formation of the sub-micron pores are not properly understood, some common features of the porous clathrate crystals can be discerned in observations of their images. A phenomenological description of porous gas hydrate formation from ice powder was developed to describe the two accessible stages of the process, they are the initial stage of the hydrate-film spreading over the ice surface and the subsequent stage of the porous hydrate layer growth limited by the clathration reaction or/and by the gas transport along the ice-hydrate interface. This theory results in the combined asymptotic relations (6) approximating the general solution (7). The reaction-limited stage should and can be distinguished from the hydrate growth controlled by the gas and water transfer through the hydrate shells surrounding the shrinking ice cores. Each rate-limiting step of the hydrate formation process and corresponding stages manifest themselves in different dependencies of the reaction degree on time. Preliminary electron microscope observations and neutron diffraction studies of the methane hydrate formation are also examined and interpreted on the basis of mathematical simulations in the companion paper by Staykova et al. (these proceedings). The theoretical predictions are in agreement with the experimental data.

### Acknowledgments

This work was supported through the BMBF Project 03G0553A in the framework of the German research initiative "Gas-Hydrate im Geosystem".